\documentclass[sigconf]{acmart}
\usepackage{multirow}
\usepackage{diagbox}
\usepackage{adjustbox}
\usepackage{subfig}
\usepackage{colortbl}
\usepackage{enumitem}
\usepackage[normalem]{ulem}
\useunder{\uline}{\ul}{}

\AtBeginDocument{%
  \providecommand\BibTeX{{%
    \normalfont B\kern-0.5em{\scshape i\kern-0.25em b}\kern-0.8em\TeX}}}

\begin{document}

\title{Fine-grained Knowledge Graph-driven Video-Language Learning for Action Recognition}


\author{Rui Zhang, Yafen Lu, Pengli Ji, Junxiao Xue, Xiaoran Yan}
\email{{ruizhang, luyf, jipengli, xuejx, yanxr}@zhejianglab.com}

\affiliation{%
  \institution{Zhejiang Lab}
  \city{Hangzhou}
  \state{Zhejiang}
  \country{China}
}


\begin{abstract}
  Recent work has explored video action recognition as a video-text matching problem and several effective methods have been proposed based on large-scale pre-trained vision-language models. However, these approaches primarily operate at a coarse-grained level without the detailed and semantic understanding of action concepts by exploiting fine-grained semantic connections between actions and body movements. To address this gap, we propose a contrastive video-language learning framework guided by a knowledge graph, termed KG-CLIP, which incorporates structured information into the CLIP model in the video domain. Specifically, we construct a multi-modal knowledge graph composed of multi-grained concepts by parsing actions based on compositional learning. By implementing a triplet encoder and deviation compensation to adaptively optimize the margin in the entity distance function, our model aims to improve alignment of entities in the knowledge graph to better suit complex relationship learning. This allows for enhanced video action recognition capabilities by accommodating nuanced associations between graph components. We comprehensively evaluate KG-CLIP on Kinetics-TPS, a large-scale action parsing dataset, demonstrating its effectiveness compared to competitive baselines. Especially, our method excels at action recognition with few sample frames or limited training data, which exhibits excellent data utilization and learning capabilities.
\end{abstract}

\ccsdesc[500]{Computing methodologies~Knowledge representation and reasoning}
\ccsdesc[500]{Computing methodologies~Activity recognition and understanding}

\keywords{Fine-grained Knowledge Graph, Video-Language Contrastive Learning, Multi-modal Learning, Action Recognition}



\maketitle

\section{Introduction}
Large-scale vision-language models (VLMs) have emerged as a dominant approach in the field of visual learning and understanding~\cite{du2023vision}. These models diverge from traditional uni-modal vision models by incorporating textual information, thereby enhancing visual understanding. This cross-modal data-based approach offers a promising solution for visual learning, and its efficacy has been substantiated~\cite{huang2021makes}.
While the inception of vision-language models was rooted in the learning of image-text pairs~\cite{radford2021learning}, recent advancements in the field have seen researchers attempting to leverage the robust multi-modal learning capabilities of these models for video learning~\cite{ju2022prompting}. Within this domain, video action recognition is one of the most sought-after tasks.

Early attempts on video action recognition mainly adopted two strategies: 2D/3D convolutional neural networks (CNNs), and two-stream networks composed of a spatial and a temporal stream~\cite{zhu2020comprehensive}. However, the complexity of these models often proved challenging when dealing with large volumes of data. Consequently, to enhance the efficiency and accuracy of video recognition models, researchers begin to explore the use of backbones generated by pre-training models for video modeling~\cite{qiu2017learning}. A pre-trained model is a saved network that has undergone training on a substantial dataset according to a unified inference objective. The data representations produced by such models often serve as highly effective features for a variety of downstream tasks. 
Currently, there is a wide array of pre-training models available. One of the most notable among these is CLIP~\cite{radford2021learning}, which employs a contrastive training scheme on 400 million image-text pairs. The large vocabulary learned by the model could promote an efficient and effective understanding of visual concepts, leading to significant improvements in various downstream tasks. As a result, CLIP-based video learning, particularly in the realm of video action recognition, has garnered considerable attention.

Recent work has explored CLIP-based video action recognition approaches, demonstrating strong performance compared to uni-modal methods, even for recognizing unseen or unfamiliar categories~\cite{DBLP:journals/corr/abs-2208-03775}. 
Primarily, CLIP acquires knowledge from both matched and unmatched relations between given pairs of images and text, of which the strategy is also adopted by the majority of CLIP-based video learning methods. However, this overemphasis on the relationship between data pairs can inadvertently lead to a degradation of scene semantics, making the model overly reliant on the co-occurrence of inputs, rather than their semantic meanings~\cite{DBLP:conf/nips/PanYHSH22}. For example, ``an action of dancing'' and ``not an action of dancing'' are opposite descriptions, but according to the co-occurrence distribution of inputs, ``not'' might be ignored, resulting in a semantic error.
Secondarily, actions are typically composite concepts. For example, ``belly dancing'' and ``salsa dancing'' both involve body movements of ``head shakes'' and ``hip turns''.
Directly mapping videos to actions without making any semantic distinction can cause misunderstandings and lead to significant cross-modal gaps~\cite{DBLP:journals/corr/abs-2303-09756}. These issues pose constraints on existing CLIP-based video action recognition models, thereby impacting their performance.

In an effort to address the aforementioned limitations, we consider to enhance semantic information in scenarios by accentuating the relationships between objects. Concurrently, we further parse actions into the discriminative language descriptions, and utilize the obtained fine-grained knowledge to bridge the gap between different modalities at a coarse-grained level~\cite{schmidt1976understanding}. To achieve this goal, we employ the technique of knowledge graph, which has been extensively researched in the field of natural language processing~\cite{DBLP:journals/eswa/ChenJX20}. Specifically, knowledge graph offers abundant information in describing the relations between entities, and the logic of triplets can also support the reasoning and tracing of knowledge within a semantic context. Therefore, by driving the vision-language model with a knowledge graph to perform action recognition tasks, we can not only enhance the semantic interaction and integration of data, but also provide a better explanation of the factors determining prediction results. This point is corroborated by existing image-language learning frameworks based on knowledge graphs~\cite{cao2022otkge,DBLP:conf/nips/PanYHSH22}. Drawing inspiration from these works, our aim is to explore a novel vision-language framework driven by knowledge graphs for action recognition in video data.

In this paper, we present KG-CLIP, a novel vision-language framework for video action recognition, which is driven by an action parsing-augmented knowledge graph. We first decompose each action into a set of fine-grained descriptions of body movements, and establish connections between videos, actions, and body movements to form a multi-modal knowledge graph, where each entity is linked by its relevant relation. Our goal is to learn triplet knowledge and infer the association between videos and actions using a vision-language framework. To this end, we adapt the CLIP model into a video-level learning architecture for multi-modal knowledge learning, and employ a Transformer as a knowledge triplet encoder to integrate the entity features based on their relations. In addition to the existing contrastive learning between vision and language modalities in CLIP, we also introduce a novel triplet learning mechanism to enable complex semantic relationship modeling and vector spatial mapping. Furthermore, considering the modality gap that emerges from projections into the joint contrastive space, we implement a deviation compensation technique to refine entity embeddings so that visual and textual representations are better aligned in this shared space. The main contributions of this paper are as follows:
\begin{itemize}[leftmargin=*]
    \item We attempt to convert video-language data into a multi-modal knowledge graph through action parsing, and propose a novel and effective multi-modal contrastive model for action recognition, leveraging the power of the knowledge graph.
    \item We carefully design a triplet learning mechanism for modeling multi-relational knowledge graph and mapping entities into a shared multi-modal space, which enables deeper understanding of complex semantic relationships and connections within the structured knowledge graph representation.  
    \item Considering the inherent challenges of joint representations across vision and language, we put forward a deviation compensation technique to actively bridge modality gaps. This allows for adaptive margin optimization over similarity measures in the contrastive space.
    \item Comprehensive benchmarking examinations demonstrate state-of-the-art performance levels compared to existing vision-language models. Notably, facilitated by the knowledge graph, our approach excels when limited training samples are available, indicating improved data utilization capabilities.
\end{itemize}


\section{Related Work}
\label{sec:rw}
In this section, we briefly review the related studies concerning Vision-Language Models (VLMs) utilized for action recognition, with a particular focus on those incorporating knowledge graphs.

\subsection{VLMs for Action Recognition}
Vision-Language Models (VLMs) have emerged as a significant research area in the field of artificial intelligence. These models aim to understand and obtain meaningful information from multi-modal data, i.e., visual (images, videos) and textual data.
Pioneering models in this domain include ViLBERT~\cite{DBLP:conf/nips/LuBPL19}, VisualBERT~\cite{DBLP:journals/corr/abs-1908-03557}, and LXMERT~\cite{DBLP:conf/emnlp/TanB19}, which leverage the Transformer architecture for joint understanding of images and text. 
The subsequently proposed CLIP (Contrastive Language-Image Pre-training) model~\cite{radford2021learning} is a significant development in the field of VLMs, which is trained on 400 million image-text pairs, and is able to return the likeliest caption or summary of an image.
Currently, many VLMs have been applied to action recognition tasks. Some famous CLIP-based approaches like ActionCLIP~\cite{DBLP:journals/corr/abs-2109-08472}, X-CLIP~\cite{DBLP:conf/eccv/NiPCZMFXL22}, EVL~\cite{DBLP:conf/eccv/LinGZGMWDQL22} and ViFi-CLIP~\cite{DBLP:journals/corr/abs-2212-03640} have all shown impressive results by processing both the visual content of a video and any associated textual descriptions to identify and classify the actions being performed. Moreover, recent advances have seen the integration of VLMs with other techniques such as attention mechanisms and temporal modeling to better capture the dynamics of actions over time~\cite{DBLP:journals/corr/abs-2304-00685}. However, despite these advancements, challenges remain in handling diverse and complex actions, particularly in environments that are too similar or too noisy.

\begin{figure*}[ht]
  \centering
  \includegraphics[width=\linewidth]{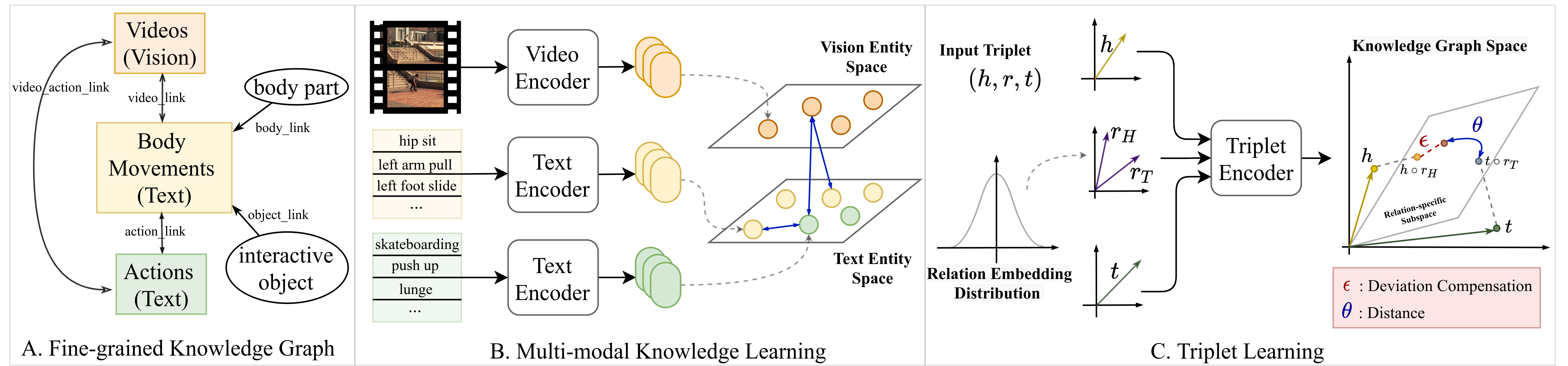}
  \caption{An overview of KG-CLIP. (A) Construction of a structured multi-modal knowledge graph via systematic action parsing to encapsulate relationships between video, text and contextual elements. (B) Encoders based on the CLIP model learn embeddings for visual and textual knowledge. The visual encoder leverages both spatial and temporal components to obtain comprehensive video representations. (C) A triplet learning module enables entity alignment by projecting head and tail entities into a relation-specific subspace, and reconciling modality gaps in the multi-modal embedding space to better facilitate action recognition.}
  \label{fig:framework}
\end{figure*}

\subsection{Multi-modal Knowledge Graph}
Multi-modal knowledge graph is an emerging field that aims to leverage multiple data modalities (e.g., structured, textual, visual) for knowledge graph learning. 
 A notable approach in this area is MMKRL~\cite{DBLP:journals/apin/LuWJHL22}, which introduces a component alignment scheme and combines it with translation methods to accomplish multi-modal knowledge representation learning. 
Another interesting work is CMGNN~\cite{DBLP:journals/tkde/FangZHWX23}, which achieves multi-modality and high-order structure modeling in an explicit and end-to-end manner under the graph neural networks with a contrastive learning framework.
Moreover, in other fields, multi-modal knowledge graphs are also widely used, such as MMUGL~\cite{DBLP:journals/corr/abs-2307-04461} in the medical field, MEduKG~\cite{DBLP:journals/information/LiSSCX22} in the education field and so on.
All of them effectively utilize multi-modal knowledge to achieve better link prediction and triplet/entity classification, which demonstrates the potential of multi-modal knowledge graph learning in various domains. 
With the impact of large-scale vision-language models, integrating multi-modal knowledge graphs has become a valuable exploration direction in current research.

\subsection{VLMs with Knowledge Graph}
Incorporating knowledge graphs into VLMs has been a recent focus in the field. \citep{zhang2024video} pioneered the integration of multi-modal structured knowledge by combining GNNs and VLMs. However, this approach is limited to ordinary graphs rather than knowledge graphs and does not explicitly learn the relationships within the graph structure.
\citep{DBLP:conf/cvpr/YeX0XY0023} discusses a DANCE strategy to enhance the commonsense reasoning ability of VLMs by leveraging commonsense knowledge graphs like ConceptNet to augment existing VLM datasets during training.
Another notable work~\cite{DBLP:conf/nips/PanYHSH22} proposes a knowledge-based pre-training framework  that semantically aligns the representations in vision and language domains by incorporating knowledge-based objectives and utilizing various knowledge graphs as training data. 
These studies demonstrate that integrating knowledge graphs into VLMs can significantly improve their performance, especially in tasks requiring commonsense reasoning or semantic understanding. However, as explorations in this direction, these methods mainly focus on the learning of images and texts, overlooking more complex data such as videos. Consequently, the development of vision-language models that leverage knowledge graphs for video understanding remains an urgent and understudied area of research.

\section{METHODOLOGY}
\label{sec:pm}
In this section, we present the proposed KG-CLIP framework for the video action recognition task. 

\subsection{Overview}
In the problem of video action recognition, given a set of videos $V = \{v_1, v_2, \ldots, v_n\}$ and a set of action labels $A = \{a_1, a_2, \ldots, a_m\}$, where each video $v_i$ can be represented as a sequence of frames $v_i = \{f_{v_i,1}, f_{v_i,2}, \ldots, f_{v_i,n_{v_i}}\}$, and the task is to map each video $v_i$ to its corresponding action label $a_j$. Related to the videos and actions is the set of body movements $B = \{b_1, b_2, \ldots, b_l\}$ that comprise each action. Formally, each action $a_j$ can be represented as a subset of body movements $\{b_{a_j,1}, b_{a_j,2}, \ldots, b_{a_j,n_{a_j}} \} \subseteq B$. Similarly, each video $v_i$ can be described by a subset of observed body movements $\{b_{v_i,1}, b_{v_i,2}, \ldots, b_{v_i,n_{v_i}} \} \subseteq B$.

We represent the knowledge using a knowledge graph $G = (O, E, R)$, where $O$ is the set of vertices representing knowledge concepts, $E$ is the set of directed edges representing prerequisite relationships, and $R=\{v\text{-}a,b\text{-}v,b\text{-}a,a\text{-}v,v\text{-}b,a\text{-}b\}$ is the set of relation types. Specifically, we define $O = \{V, A, B\}$ and represent a triplet as $(h, r, t) \in E$ where $h,t \in O$ are vertices and $r \in R$ indicates the prerequisite relation type. The knowledge graph $G$ captures the relationships among videos, actions and body movements.

Figure~\ref{fig:framework} provides an overview of the proposed KG-CLIP framework for contrastive video-language learning guided by a multi-modal knowledge graph. Specifically, we leverage the pre-trained multi-modal encoders from CLIP to encode visual and textual data into modality-specific embedding spaces, with the addition of a temporal encoder on top of the static image encoder to obtain video representations. The learned embeddings initialize entity representations within the knowledge graph. Knowledge graph triplets are then concatenated and encoded to construct a joint relation and entity space. Head and tail entities are projected into relation-specific subspaces, followed by a cross-modal deviation compensation technique to reconcile modality gaps and promote similarity measures for the task of action recognition. This framework couples contrastive video-language learning with structured knowledge graph representations to enable joint modeling and understanding of knowledge concepts and relationships.

\subsection{Fine-grained Knowledge Graph}
The constructed knowledge graph $G$, shown in Figure~\ref{fig:framework}, is a critical component of our proposed multi-modal knowledge learning framework for video action recognition. The knowledge graph contains three primary node types, i.e., videos, actions and body movements. Each body movement node represents a composite concept consisting of the corresponding body part, part state, and interacted object. For simplicity, we represent each body movement as a single node with a descriptive text label. The edges in the knowledge graph capture the semantic relationships between the nodes. For example, the relationship between videos and actions may be "performing," "executing," or "depicting an action of." 
In order to define this rich semantic relationship in $G$, we generally refer to this kind of relation as "video\_action\_link". 
In practical training, we consider the asymmetric bidirectional relations between knowledge and encode both of them for multi-modal learning, of which the result serve as a vital input to the subsequent knowledge fusion stage of our approach.

\subsection{Multi-modal Knowledge Learning}
In the multi-modal knowledge learning stage, we first extract frames from the input video clips and apply a video encoder comprising the CLIP image encoder $ImgEnc$ and a mean pooling component $Tem$ for temporal encoding to obtain video representations $\mathbf{X}^V \in \mathbb{R}^{|V| \times d}$, where $d$ is the feature dimension.
For textual knowledge, we directly utilize the pre-trained CLIP text encoder $TxtEnc$ to generate embeddings for actions and body movements, denoted as $\mathbf{X}^A \in \mathbb{R}^{|A| \times d}$ and $\mathbf{X}^B \in \mathbb{R}^{|B| \times d}$ respectively. Here, we adopt the ViT-B architecture for CLIP, consisting of a 12-layer image encoder and 12-layer 512-wide text encoder with 8 attention heads.
In this manner, multi-modal knowledge is encoded into modality-specific embedding spaces, serving to initialize entity representations for subsequent knowledge graph learning.

We also implement multi-modal contrastive learning to optimize the encoders at the modal encoding step prior to knowledge graph learning. Unlike the learning strategy of the original CLIP model which only contrasts vision and language modalities, we learn associations within the multi-modal knowledge graph by reducing differences between videos and body movements, actions and body movements, and videos and actions during encoding. This lays the foundation for subsequent graph learning.
We define the Kullback–Leibler (KL) divergence to maximize agreement between positively paired instances while minimizing agreement between negatively paired instances. Taking a batch of video-action pairs $\mathcal{T}$ as an example, the contrastive loss is formulated as:
\begin{equation}
\label{eq:mm_loss}
\begin{split}
    \mathcal{L}^{mm}_{v-a}&=\frac{1}{2} \mathbb{E}_{({v}, {a}) \sim \mathcal{T}}\left[\mathcal{D}_{KL}\left(\mathcal{P}(a|v) || {q}\right)+\mathcal{D}_{KL}\left(\mathcal{P}({v|a}) || {q}^{'}\right)\right]
\end{split}
\end{equation}
where $q$ is the ground-truth, and $\mathcal{P}$ indicates the distribution which can be approximated by the similarity scores as follows:
\begin{equation}
\label{eq:kl_p}
\begin{split}
    \mathcal{P}(a|v) &= \frac{\exp(cos(\mathbf{X}^v,\mathbf{X}^a))}{\sum_{(v, *) \in \mathcal{T}}\exp(cos(\mathbf{X}^v,\mathbf{X}^{*}))} \\
    \mathcal{P}(v|a) &= \frac{\exp(cos(\mathbf{X}^a,\mathbf{X}^v))}{\sum_{(*, a) \in \mathcal{T}}\exp(cos(\mathbf{X}^a,\mathbf{X}^{*}))} \\
\end{split}
\end{equation}

\subsection{Triplet Learning}
The representations from the modal encoders as the initialization vector of entities in the knowledge graph, serve the triplet learning module. The purpose of this module is to conduct structured learning of the knowledge graph and perform triplet computations in an optimal shared space.
Notably, considering the diverse semantic relationships between knowledge concepts, and the utility of these relations as prompt guidance in multi-modal modeling, we opt not to use fixed linguistic text to represent relations. Instead, we introduce a learnable relation matrix to flexibly associate knowledge and facilitate triplet learning.
Inspired by PairRE~\cite{DBLP:conf/acl/ChaoHWC20}, we adopt paired relation vectors, suitable for modeling complex and multi-pattern relationships. Unpon that, we define the relation matrix $\mathbf{X}_R \in \mathbb{R}^{|R| \times 2d}$ from a Gaussian distribution.

{\bf Triplet Encoding.}
In this module, a triplet encoder $TriEnc$ is first applied to model a joint distribution of entities and relations by learning the concatenated sequence representation of each triplet. 
Specifically, given a triplet $(h,r,t)$, the initial embeddings of its elements correspond to $\mathbf{X}_h \in \mathbb{R}^{d}$, $\mathbf{X}_r \in \mathbb{R}^{2d}$ and $\mathbf{X}_t \in \mathbb{R}^{d}$ respectively.
According to the paired relation vector operation, the relation vector $\mathbf{X}r$ can be split into paired chunks $\mathbf{X}{r^h} \in \mathbb{R}^{d}$ and $\mathbf{X}{r^t} \in \mathbb{R}^{d}$. Therefore, the concatenated input sequence can be expressed as $[\mathbf{X}_h,\mathbf{X}_{r^h},\mathbf{X}_{r^t},\mathbf{X}_t]^\intercal$, and the triplet encoding process is formulated as:
\begin{equation}
\label{eq:trienc}
\left[\begin{array}{c}
\mathbf{Z}_h \\
\;\mathbf{Z}_{r^h} \\
\;\mathbf{Z}_{r^t}\\
\mathbf{Z}_t
\end{array}\right]
= TriEnc \left(
\left[\begin{array}{c}
\mathbf{X}_h \\
\;\mathbf{X}_{r^h} \\
\;\mathbf{X}_{r^t}\\
\mathbf{X}_t
\end{array}\right]+\left[\begin{array}{c}
\mathbf{P}_1 \\
\mathbf{P}_{2} \\
\mathbf{P}_{3}\\
\mathbf{P}_4
\end{array}\right] \right) + 
\left[\begin{array}{c}
\mathbf{X}_h \\
\;\mathbf{X}_{r^h} \\
\;\mathbf{X}_{r^t}\\
\mathbf{X}_t
\end{array}\right]
\end{equation}
where $\mathbf{P}$ is the trainable positional embedding matrix to emphasize the status of elements in the sequence during triplet encoding, and the triplet encoder $TriEnc$ follows the traditional architecture of a 3-layer Transformer. 

{\bf Entity Projection.}
After triplet encoding, entities and relations occupy a joint vector space. The paired relation vector strategy allows an entity to have distributed representations specific to different relations. This makes it easy to adaptively adjust the loss margin, which can alleviate issues in modeling complex relations. Leveraging this, we use the two pairwise relation vectors to project the head and tail entities into relation-specific Euclidean subspaces, formulated as:
\begin{equation}
\label{eq:entproj}
\begin{split}
    \mathbf{Z}_{h}^{'} = \mathbf{Z}_{h} \circ  \mathbf{Z}_{r^h}\quad ,\quad
    \mathbf{Z}_{t}^{'} = \mathbf{Z}_{t} \circ  \mathbf{Z}_{r^t}
\end{split}
\end{equation}
where $\circ$ denotes entry-wise Hadamard product. For valid triplets, the projected head and tail vectors should exhibit proximity, while unrelated heads and tails should remain distant. Thus, the distance between $\mathbf{Z}_{h}^{'}$ and $\mathbf{Z}_{t}^{'}$ indicates triplet plausibility, thereby enhancing structural understanding of the knowledge graph.

\begin{figure}
    \centering
    \includegraphics[width=0.95\linewidth]{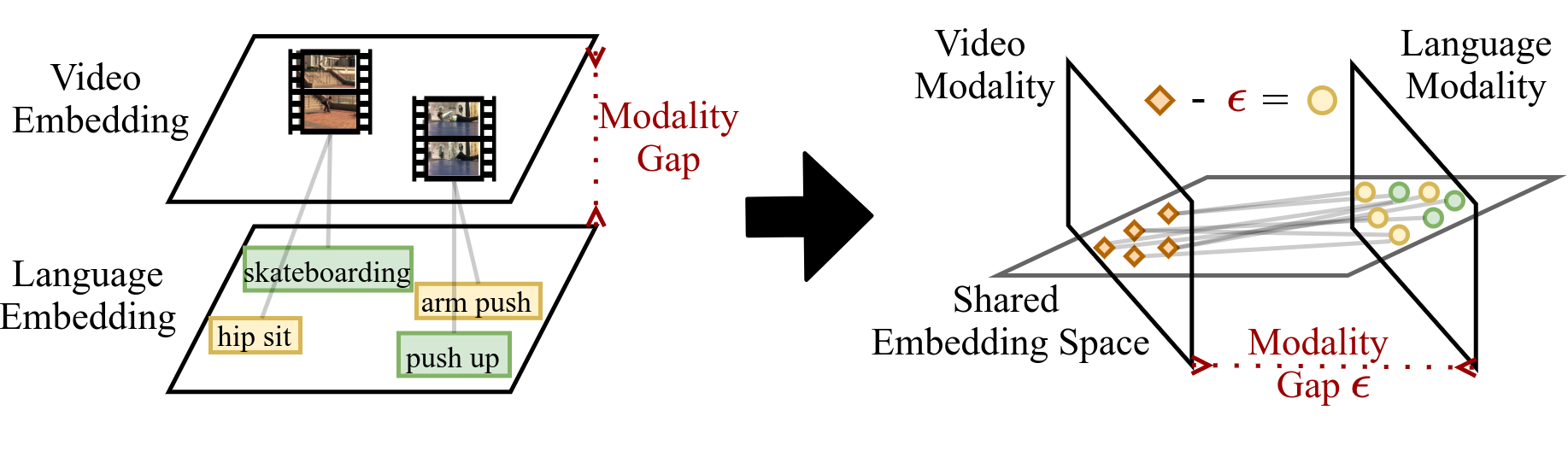}
    \caption{An illustration of modality gap in video-language contrastive learning.}
    \label{fig:modality_gap}
\end{figure}

{\bf Deviation Compensation.}
Recent work~\cite{liang2022mind} shows that different modalities are embedded at a certain distance in their shared representation in multi-modal models, known as the modality gap.
This gap remains persistent even under various downstream tasks, as shown in Figure~\ref{fig:modality_gap}. In our task, where entity vectors from the knowledge graph are derived from the multi-modal encoders, we believe that this gap persists throughout multi-modal learning and triplet learning, ultimately manifesting as a gap between the head and tail entities within each triplet. 

Typically, the modality gap is defined as the disparity between embedding centers of different modalities. 
However, we take into account the changes in the knowledge vector space during the learning process, allowing a simply yet more effective and adaptable solution. Specifically, we introduce a deviation compensation technique that utilizes a learnable vector $\epsilon \in \mathbb{R}^d$ within relation-specific Euclidean subspaces to mitigate the gap and improve performance on entity alignment, formalized as:
\begin{equation}
\label{eq:dc}
    \bar{\mathbf{Z}}^{''}_{h} =  \bar{\mathbf{Z}}^{'}_{h} - \epsilon
\end{equation}
where $\bar{\mathbf{Z}}^{'}_{h}$ and $\bar{\mathbf{Z}}^{''}_{h}$ are the embedding centers before and after deviation compensation.
By modifying the gap between modalities through this simple linear calculation, entity representations are better optimized for triplet computations to improve learning of knowledge graph structure.

{\bf Training Objective}
In the relation-specific subspaces, triplet rationality is evaluated by the distance between the projections of the head and tail entities. Accordingly, the optimization objective for this triplet learning module is to minimize the distance between the head embedding, adjusted by the deviation compensation, and the tail embedding, for all valid triplets. 
Given the knowledge graph under study incorporates a substantial number of intricate relations, outliers may distort the embedding space and impede model optimization. Therefore, we opt for the cosine distance rather than absolute or squared Euclidean measures (conventionally employed in knowledge graph representation learning) to mitigate the influence of potential outliers.
The formula expressing the distance calculation is as follows:
\begin{equation}
    \theta = \cos(\mathbf{Z}_h^{''},\mathbf{Z}_t^{'}) = \dfrac{\mathbf{Z}_h^{''} \cdot \mathbf{Z}_t^{'}}{\lVert \mathbf{Z}_h^{''} \rVert \lVert \mathbf{Z}_t^{'} \rVert}
    \label{eq:cosine}
\end{equation}

Another vital component of this module is the concurrent training on both forward triplets and their reverse counterparts during each learning iteration. Specifically, the optimization objective for every training cycle is defined as the aggregate of distance losses computed from the input triplet $(h,r,t)$ that yields distance $\theta$, along with its reverse triplet $(t,r^{'}, h)$ that yields distance $\theta^{'}$. By simultaneously minimizing discrepancies in both the forward and backward directions, the model is endowed with a symmetrized inductive bias better suited for aligning entities and relations across the knowledge graph.

Despite aggregating both forward and reverse triplet losses and employing cosine distance, directly optimizing this alone still requires sampling numerous negative examples. To circumvent this computational burden, we adopt KL divergence between the similarity distribution and the ground-truth distribution for the optimization objective of the triplet learning module. By concurrently minimizing this global loss, the model converges more efficiently by considering violation cases beyond the limited triplet samples in each batch. The formula is expressed as follows:
\begin{equation}
\label{eq:tri_loss_1}
\begin{split}
\mathcal{L}^{tri}_{h-t} = \frac{1}{2}\mathbb{E}_{({h}, {t}) \sim \mathcal{T}^{'}} [\mathcal{D}_{KL}(\mathcal{P}(t|h)||\mathcal{Q}) + \mathcal{D}_{KL}(\mathcal{P}(h|t)||\mathcal{Q}^{'})]
\end{split}
\end{equation}
where $\mathcal{T}^{'}$ denotes a batch of triplets, $\mathcal{Q}$ and $\mathcal{Q}^{'}$ represent the ground-truth similarity scores for the forward and reverse directions respectively, and the computation of the similarity distribution $\mathcal{P}$ adheres to Equation~\ref{eq:kl_p}, shown below:
\begin{equation}
\begin{split}
    \mathcal{P}(t|h) &= \frac{\exp(\theta)}{\sum_{(h, *) \in \mathcal{T}^{'}}\exp(cos(\mathbf{Z}_{h}^{''},\mathbf{Z}_{*}^{'}))} \\
    \mathcal{P}(h|t) &= \frac{\exp(\theta^{'})}{\sum_{(t, *) \in \mathcal{T}^{'}}\exp(cos(\mathbf{Z}_{t}^{''},\mathbf{Z}_{*}^{'}))}
\end{split}
\end{equation}

Integrating the loss from the triplet learning module with that from the prior multi-modal learning module, the overall optimization objective for the proposed KG-CLIP framework can be formulated as:
\begin{equation}
\begin{split}
    \mathcal{L} = \frac{1}{N} \sum_{* \in \mathcal{E}}\left(\mathcal{L}^{tri}_{*} + \lambda \mathcal{L}^{mm}_{*} \right)
\end{split}
\label{eq:final_loss}
\end{equation}
where $N$ is the number of batch samples, $\mathcal{E}=\{v\text{-}a,b\text{-}v,b\text{-}a\}$ is the sets of knowledge relation types, and $\lambda$ is the coefficient of the multi-modal loss.

\subsection{Action Inference}
In multi-modal learning, recognition prediction relies on a similarity matrix as the final output to determine correlations between vision and text. For the proposed KG-CLIP framework, the culminating similarity matrix is derived through the strategic assimilation of the previously elucidated similarity computations, formulated as follows:
\begin{equation}
\begin{split}
    \mathbf{S}^{output} = \frac{1}{2}(\mathbf{S}^{mm} + \mathbf{S}^{tri})
\end{split}
\end{equation}
where $\mathbf{S}^{mm} = cos(\mathbf{X}^v,\mathbf{X}^a)$ signifies the similarity matrix derived within the multi-modal learning component, and $\mathbf{S}^{tri}$ represents that obtained from the triplet learning module, formally defined as:
\begin{equation}
\begin{split}
    {S}^{tri} = \frac{1}{2}(cos(\mathbf{Z}_{v}^{''},\mathbf{Z}_{a}^{'}) + cos(\mathbf{Z}_{a}^{''},\mathbf{Z}_{v}^{'})) 
\end{split}
\end{equation}

\begin{table}[t]
\caption{Statistics of the multi-modal knowledge graph.}
\begin{tabular}{|cc|cc|}
\hline
\multicolumn{2}{|c|}{\textbf{Node}}                                                                                            & \multicolumn{2}{c|}{\textbf{Edge}}  \\ \hline
\multicolumn{1}{|c|}{\textbf{Types}}                                                                            & \textbf{Number}                & \multicolumn{1}{c|}{\textbf{Types}} & \textbf{Number} \\ \hline \hline
\multicolumn{1}{|c|}{\multirow{2}{*}{Action (text)}}                                                   & \multirow{2}{*}{24}   & \multicolumn{1}{c|}{v-b}   & 5040   \\ \cline{3-4} 
\multicolumn{1}{|c|}{}                                                                                 &                       & \multicolumn{1}{c|}{b-v}   & 5040   \\ \hline
\multicolumn{1}{|c|}{\multirow{2}{*}{\begin{tabular}[c]{@{}c@{}}Body \\ Movement (text)\end{tabular}}} & \multirow{2}{*}{391}  & \multicolumn{1}{c|}{a-b}   & 801    \\ \cline{3-4} 
\multicolumn{1}{|c|}{}                                                                                 &                       & \multicolumn{1}{c|}{b-a}   & 801    \\ \hline
\multicolumn{1}{|c|}{\multirow{2}{*}{Video (vision)}}                                                  & \multirow{2}{*}{3809} & \multicolumn{1}{c|}{v-a}   & 3809   \\ \cline{3-4} 
\multicolumn{1}{|c|}{}                                                                                 &                       & \multicolumn{1}{c|}{a-v}   & 3809   \\ \hline
\end{tabular}
\label{tb:stats}
\end{table}
\begin{table*}[t]
\centering
\caption{Overall results (\%) of our proposed KG-CLIP on four Kinetics-TPS datasets with the previous baselines. Bold indicates the best performance and underline indicates the runner-up.}
\resizebox{0.98\linewidth}{!}{%
\begin{tabular}{|ccccccccccccc|}
\hline
\multicolumn{1}{|c|}{\textbf{Backbones}}              & \multicolumn{6}{c|}{\textbf{ViT-B/32}}       & \multicolumn{6}{c|}{\textbf{ViT-B/16}}                               \\ \hline
\multicolumn{1}{|c|}{\textbf{\# frames}}              & \multicolumn{2}{c|}{\textbf{4}}                                                               & \multicolumn{2}{c|}{\textbf{8}}                                                               & \multicolumn{2}{c|}{\textbf{16}}                                                              & \multicolumn{2}{c|}{\textbf{4}}                                                                                           & \multicolumn{2}{c|}{\textbf{8}}                                                                                           & \multicolumn{2}{c|}{\textbf{16}}                                                                                          \\ \hline
\multicolumn{1}{|c|}{\diagbox[dir=SE]{\textbf{Methods}}{\textbf{Accuracy}}}                       & \multicolumn{1}{c|}{\textbf{Top-1}}           & \multicolumn{1}{c|}{\textbf{Top-5}}           & \multicolumn{1}{c|}{\textbf{Top-1}}           & \multicolumn{1}{c|}{\textbf{Top-5}}           & \multicolumn{1}{c|}{\textbf{Top-1}}           & \multicolumn{1}{c|}{\textbf{Top-5}}           & \multicolumn{1}{c|}{\textbf{Top-1}}                         & \multicolumn{1}{c|}{\textbf{Top-5}}                         & \multicolumn{1}{c|}{\textbf{Top-1}}                         & \multicolumn{1}{c|}{\textbf{Top-5}}                         & \multicolumn{1}{c|}{\textbf{Top-1}}                         & \textbf{Top-5}                                              \\ \hline\hline
\multicolumn{13}{|c|}{training set : test set = 2 : 8}     \\ \hline

\multicolumn{1}{|c|}{ActionCLIP}                      & \multicolumn{1}{c|}{78.71}                                  & \multicolumn{1}{c|}{98.29}                                  & \multicolumn{1}{c|}{79.63}                                  & \multicolumn{1}{c|}{{\ul 98.85}}                            & \multicolumn{1}{c|}{79.66}                                  & \multicolumn{1}{c|}{{\ul 98.52}}                            & \multicolumn{1}{c|}{82.51}                                  & \multicolumn{1}{c|}{98.43}                                  & \multicolumn{1}{c|}{83.07}                                  & \multicolumn{1}{c|}{98.59}                                  & \multicolumn{1}{c|}{83.89}                                  & 98.10          \\ 
\multicolumn{1}{|c|}{ViFi-CLIP}                       & \multicolumn{1}{c|}{78.51}                                  & \multicolumn{1}{c|}{98.16}                                  & \multicolumn{1}{c|}{78.87}                                  & \multicolumn{1}{c|}{97.74}                                  & \multicolumn{1}{c|}{79.27}                                  & \multicolumn{1}{c|}{97.54}                                  & \multicolumn{1}{c|}{{\ul 82.78}}                            & \multicolumn{1}{c|}{{\ul 98.95}}                         & \multicolumn{1}{c|}{{\ul 83.33}}                            & \multicolumn{1}{c|}{98.10}                                  & \multicolumn{1}{c|}{{\ul 84.06}}                            & {\ul 98.20}    \\ 
\multicolumn{1}{|c|}{X-CLIP}                          & \multicolumn{1}{c|}{{\ul 80.35}}                            & \multicolumn{1}{c|}{{\ul 98.36}}                            & \multicolumn{1}{c|}{{\ul 80.22}}                            & \multicolumn{1}{c|}{98.49}                                  & \multicolumn{1}{c|}{{\ul 79.99}}                            & \multicolumn{1}{c|}{98.00}                                  & \multicolumn{1}{c|}{82.32}                                  & \multicolumn{1}{c|}{98.66}                                  & \multicolumn{1}{c|}{83.24}                                  & \multicolumn{1}{c|}{{\ul 98.72}}                            & \multicolumn{1}{c|}{83.43}                                  & {\ul 98.20}    \\ 
\rowcolor[HTML]{EFEFEF} 
\multicolumn{1}{|c|}{\cellcolor[HTML]{EFEFEF}KG-CLIP} & \multicolumn{1}{c|}{\cellcolor[HTML]{EFEFEF}\textbf{83.14}} & \multicolumn{1}{c|}{\cellcolor[HTML]{EFEFEF}\textbf{99.21}} & \multicolumn{1}{c|}{\cellcolor[HTML]{EFEFEF}\textbf{83.66}} & \multicolumn{1}{c|}{\cellcolor[HTML]{EFEFEF}\textbf{99.25}} & \multicolumn{1}{c|}{\cellcolor[HTML]{EFEFEF}\textbf{84.12}} & \multicolumn{1}{c|}{\cellcolor[HTML]{EFEFEF}\textbf{99.21}} & \multicolumn{1}{c|}{\cellcolor[HTML]{EFEFEF}\textbf{85.40}} & \multicolumn{1}{c|}{\cellcolor[HTML]{EFEFEF}\textbf{99.02}}    & \multicolumn{1}{c|}{\cellcolor[HTML]{EFEFEF}\textbf{85.63}} & \multicolumn{1}{c|}{\cellcolor[HTML]{EFEFEF}\textbf{99.34}} & \multicolumn{1}{c|}{\cellcolor[HTML]{EFEFEF}\textbf{87.14}} & \multicolumn{1}{c|}{\cellcolor[HTML]{EFEFEF}\textbf{99.15}}                                           \\ \hline\hline

\multicolumn{13}{|c|}{training set : test set = 4 : 6}                                                                                             \\ \hline
\multicolumn{1}{|c|}{ActionCLIP}                      & \multicolumn{1}{c|}{{\ul 83.60}}                    & \multicolumn{1}{c|}{98.95}                    & \multicolumn{1}{c|}{83.81}                    & \multicolumn{1}{c|}{98.73}                    & \multicolumn{1}{c|}{84.69}                    & \multicolumn{1}{c|}{98.78}                    & \multicolumn{1}{c|}{86.09}                                  & \multicolumn{1}{c|}{{\ul 99.30}}                            & \multicolumn{1}{c|}{86.83}                                  & \multicolumn{1}{c|}{99.04}                                  & \multicolumn{1}{c|}{88.41}                                  &  99.30                                                 \\ 
\multicolumn{1}{|c|}{ViFi-CLIP}                       & \multicolumn{1}{c|}{{\ul 83.60}}                    & \multicolumn{1}{c|}{{\ul 99.16}}                    & \multicolumn{1}{c|}{83.68}                    & \multicolumn{1}{c|}{99.04}                    & \multicolumn{1}{c|}{{\ul 84.86}}                    & \multicolumn{1}{c|}{98.91}                    & \multicolumn{1}{c|}{86.05}                                  & \multicolumn{1}{c|}{99.21}                                  & \multicolumn{1}{c|}{{\ul 87.66}}                            & \multicolumn{1}{c|}{{\ul 99.13}}                            & \multicolumn{1}{c|}{{\ul 88.45}}                            & 99.30                                                \\ 
\multicolumn{1}{|c|}{X-CLIP}                          & \multicolumn{1}{c|}{82.90}                    & \multicolumn{1}{c|}{98.91}                    & \multicolumn{1}{c|}{{\ul 84.51}}                    & \multicolumn{1}{c|}{{\ul 99.17}}                    & \multicolumn{1}{c|}{83.82}                    & \multicolumn{1}{c|}{{\ul 99.13}}                    & \multicolumn{1}{c|}{{\ul 86.57}}                            & \multicolumn{1}{c|}{99.21}                                  & \multicolumn{1}{c|}{86.83}                                  & \multicolumn{1}{c|}{{\ul 99.13}}                            & \multicolumn{1}{c|}{87.84}                                  & {\ul 99.56}                                              \\ 
\rowcolor[HTML]{EFEFEF} 
\multicolumn{1}{|c|}{\cellcolor[HTML]{EFEFEF}KG-CLIP} & \multicolumn{1}{c|}{\cellcolor[HTML]{EFEFEF}\textbf{86.26}} & \multicolumn{1}{c|}{\cellcolor[HTML]{EFEFEF}\textbf{99.39}} & \multicolumn{1}{c|}{\cellcolor[HTML]{EFEFEF}\textbf{87.58}} & \multicolumn{1}{c|}{\cellcolor[HTML]{EFEFEF}\textbf{99.52}} & \multicolumn{1}{c|}{\cellcolor[HTML]{EFEFEF}\textbf{87.49}} & \multicolumn{1}{c|}{\cellcolor[HTML]{EFEFEF}\textbf{99.48}} & \multicolumn{1}{c|}{\cellcolor[HTML]{EFEFEF}\textbf{87.84}} & \multicolumn{1}{c|}{\cellcolor[HTML]{EFEFEF}\textbf{99.48}} & \multicolumn{1}{c|}{\cellcolor[HTML]{EFEFEF}\textbf{89.72}} & \multicolumn{1}{c|}{\cellcolor[HTML]{EFEFEF}\textbf{99.48}} & \multicolumn{1}{c|}{\cellcolor[HTML]{EFEFEF}\textbf{89.81}} & \multicolumn{1}{c|}{\cellcolor[HTML]{EFEFEF}\textbf{99.78}}          \\ \hline\hline

\multicolumn{13}{|c|}{training set : test set = 6 : 4}                                    \\ \hline
\multicolumn{1}{|c|}{ActionCLIP}              
& \multicolumn{1}{c|}{{\ul 84.91}}                    
& \multicolumn{1}{c|}{{\ul 99.02}}                   
& \multicolumn{1}{c|}{85.89}                    
& \multicolumn{1}{c|}{{\ul 99.28}}                    
& \multicolumn{1}{c|}{85.96}                    
& \multicolumn{1}{c|}{{\ul 98.95}}                    
& \multicolumn{1}{c|}{{\ul 88.45}}                           
& \multicolumn{1}{c|}{{\ul 99.21}}        
& \multicolumn{1}{c|}{{\ul 89.57}}                            
& \multicolumn{1}{c|}{99.34}         
& \multicolumn{1}{c|}{89.89}                                  
& 99.28                                                      
\\ 
\multicolumn{1}{|c|}{ViFi-CLIP}               
& \multicolumn{1}{c|}{84.58}                    
& \multicolumn{1}{c|}{98.75}                    
& \multicolumn{1}{c|}{85.17}                    
& \multicolumn{1}{c|}{98.56}                    
& \multicolumn{1}{c|}{85.89}                   
& \multicolumn{1}{c|}{98.89}                   
& \multicolumn{1}{c|}{87.73}                                  
& \multicolumn{1}{c|}{99.02}                                  
& \multicolumn{1}{c|}{{\ul 89.57}}                           
& \multicolumn{1}{c|}{{\ul 99.41}}                         
& \multicolumn{1}{c|}{{\ul 90.09}}                           
& {\ul 99.41}                                             
\\ 
\multicolumn{1}{|c|}{X-CLIP}                          
& \multicolumn{1}{c|}{83.73}                    
& \multicolumn{1}{c|}{98.43}                    
& \multicolumn{1}{c|}{{\ul 86.16}}                    
& \multicolumn{1}{c|}{ 99.08}                   
& \multicolumn{1}{c|}{{\ul 86.48}}                    
& \multicolumn{1}{c|}{98.69}                    
& \multicolumn{1}{c|}{87.21}                                  
& \multicolumn{1}{c|}{99.08}                            
& \multicolumn{1}{c|}{89.30}                                  
& \multicolumn{1}{c|}{99.08}                                  
& \multicolumn{1}{c|}{89.83}                                  
&  { 99.34}                                              \\ 
\rowcolor[HTML]{EFEFEF} 
\multicolumn{1}{|c|}{\cellcolor[HTML]{EFEFEF}KG-CLIP} & \multicolumn{1}{c|}{\cellcolor[HTML]{EFEFEF}\textbf{87.14}} & \multicolumn{1}{c|}{\cellcolor[HTML]{EFEFEF}\textbf{99.41}} & \multicolumn{1}{c|}{\cellcolor[HTML]{EFEFEF}\textbf{87.47}} & \multicolumn{1}{c|}{\cellcolor[HTML]{EFEFEF}\textbf{99.41}} & \multicolumn{1}{c|}{\cellcolor[HTML]{EFEFEF}\textbf{87.66}} & \multicolumn{1}{c|}{\cellcolor[HTML]{EFEFEF}\textbf{99.21}} & \multicolumn{1}{c|}{\cellcolor[HTML]{EFEFEF}\textbf{89.57}} & \multicolumn{1}{c|}{\cellcolor[HTML]{EFEFEF}\textbf{99.34}} & \multicolumn{1}{c|}{\cellcolor[HTML]{EFEFEF}\textbf{90.09}} & \multicolumn{1}{c|}{\cellcolor[HTML]{EFEFEF}\textbf{99.41}} & \multicolumn{1}{c|}{\cellcolor[HTML]{EFEFEF}\textbf{90.62}} & \multicolumn{1}{c|}{\cellcolor[HTML]{EFEFEF}\textbf{99.48}} \\ \hline\hline
\multicolumn{13}{|c|}{training set : test set = 8 : 2   }    \\ \hline
\multicolumn{1}{|c|}{ActionCLIP}              
& \multicolumn{1}{c|}{{\ul 86.48}}                    
& \multicolumn{1}{c|}{{\ul 98.82}}                    
& \multicolumn{1}{c|}{86.88}                    
& \multicolumn{1}{c|}{98.82}                    
& \multicolumn{1}{c|}{{\ul 87.66}}                    
& \multicolumn{1}{c|}{{\ul 99.21}}                    
& \multicolumn{1}{c|}{89.37}                                  
& \multicolumn{1}{c|}{99.02}                  
& \multicolumn{1}{c|}{91.21}                                  
& \multicolumn{1}{c|}{99.34}                 
& \multicolumn{1}{c|}{91.08}                                  
& {\ul 99.61}                                               
\\ 
\multicolumn{1}{|c|}{ViFi-CLIP}               
& \multicolumn{1}{c|}{85.96}                    
& \multicolumn{1}{c|}{\textbf{99.21}}                   
& \multicolumn{1}{c|}{{\ul 87.80}}                    
& \multicolumn{1}{c|}{{\ul 98.95}}                    
& \multicolumn{1}{c|}{86.61}                    
& \multicolumn{1}{c|}{{99.08}}                    
& \multicolumn{1}{c|}{88.71}                                  
& \multicolumn{1}{c|}{99.21}                  
& \multicolumn{1}{c|}{89.90}                                  
& \multicolumn{1}{c|}{{\ul 99.48}}            
& \multicolumn{1}{c|}{90.03}                                  
& 99.34                                                       
\\ 
\multicolumn{1}{|c|}{X-CLIP}                  
& \multicolumn{1}{c|}{{\ul 86.48}}                    
& \multicolumn{1}{c|}{98.43}                    
& \multicolumn{1}{c|}{87.53}                    
& \multicolumn{1}{c|}{98.82}                    
& \multicolumn{1}{c|}{86.61}                    
& \multicolumn{1}{c|}{98.69}                    
& \multicolumn{1}{c|}{{\ul 89.63}}                            
& \multicolumn{1}{c|}{{\ul 99.34}}            
& \multicolumn{1}{c|}{{\ul 91.60}}                            
& \multicolumn{1}{c|}{{\ul 99.48}}            
& \multicolumn{1}{c|}{{\ul 91.60}}                            
& \textbf{99.73}                                           
\\ 
\rowcolor[HTML]{EFEFEF} 
\multicolumn{1}{|c|}{\cellcolor[HTML]{EFEFEF}KG-CLIP} 
& \multicolumn{1}{c|}{\cellcolor[HTML]{EFEFEF}\textbf{88.98}} 
& \multicolumn{1}{c|}{\cellcolor[HTML]{EFEFEF}\textbf{99.21}} 
& \multicolumn{1}{c|}{\cellcolor[HTML]{EFEFEF}\textbf{88.19}} 
& \multicolumn{1}{c|}{\cellcolor[HTML]{EFEFEF}\textbf{99.21}} 
& \multicolumn{1}{c|}{\cellcolor[HTML]{EFEFEF}\textbf{88.85}} 
& \multicolumn{1}{c|}{\cellcolor[HTML]{EFEFEF}\textbf{99.34}} 
& \multicolumn{1}{c|}{\cellcolor[HTML]{EFEFEF}\textbf{91.47}} 
& \multicolumn{1}{c|}{\cellcolor[HTML]{EFEFEF}\textbf{99.48}} 
& \multicolumn{1}{c|}{\cellcolor[HTML]{EFEFEF}\textbf{92.52}} 
& \multicolumn{1}{c|}{\cellcolor[HTML]{EFEFEF}\textbf{99.48}} 
& \multicolumn{1}{c|}{\cellcolor[HTML]{EFEFEF}\textbf{92.52}} 
& \multicolumn{1}{c|}{\cellcolor[HTML]{EFEFEF}{\ul 99.61}}                                                      \\ \hline
\end{tabular}
}
\label{tab:overall}
\end{table*}
\section{EXPERIMENTS}
\label{sec:exp}
We report extensive experiments designed to evaluate the performance of KG-CLIP in this section.

\subsection{Experimental Setup}
\noindent \textbf{Dataset.} We conduct experiments on the Kinetics-TPS dataset\footnote{https://deeperaction.github.io/}, which parses human actions through compositional learning of body part movements. This dataset provides detailed human part annotations, including 10 different body parts, 74 distinct part states, and 75 categories of interactive objects, to enable in-depth video action understanding. Kinetics-TPS contains 3,809 publicly available videos spanning 24 complex human action categories in unconstrained environments, selected from the larger Kinetics-700 dataset. For the multi-modal knowledge graph constructed in this work, statistics are presented in Table~\ref{tb:stats}.

\noindent\textbf{Baselines.} 
As our proposed method represents the first work on CLIP-based video action recognition guided by a multi-modal knowledge graph, we compare against several recent CLIP-based multi-modal algorithms: 
\begin{itemize}[leftmargin=*]
    \item ActionCLIP\footnote{https://github.com/sallymmx/ActionCLIP}~\cite{DBLP:journals/corr/abs-2109-08472} is one of the first works to adapt CLIP for video action recognition, which proposes a "pre-train, prompt and finetune" paradigm.
    \item ViFi-CLIP\footnote{https://github.com/muzairkhattak/ViFi-CLIP}~\cite{DBLP:journals/corr/abs-2212-03640} proposes a solution about how to adapt the image-based CLIP model to the video domain by fine-tuning the CLIP model on videos without adding any new modules or components.
    \item X-CLIP\footnote{https://github.com/microsoft/VideoX/tree/master/X-CLIP}~\cite{DBLP:conf/eccv/NiPCZMFXL22} adapts the CLIP to the video domain without pre-training a new model from scratch and proposes a cross-frame attention unit to capture the temporal information and inter-object relationships in videos.
\end{itemize}

\noindent \textbf{Evaluation Metrics.}
To evaluate the video recognition performance of our proposed model, we use Top-1 accuracy (Top-1) and Top-5 accuracy (Top-5) as the recognition metrics, following conventions in previous action recognition work~\cite{DBLP:journals/corr/abs-2109-08472,DBLP:journals/corr/abs-2303-09756,DBLP:conf/eccv/LinGZGMWDQL22,DBLP:conf/eccv/NiPCZMFXL22,DBLP:journals/corr/abs-2212-03640}. Top-N accuracy refers to the proportion of test samples for which the ground truth action is predicted within the top-N highest similarity scores.

\noindent \textbf{Experimental Settings.}
We implement the experiments using PyTorch~\cite{pytorch} with the AdamW optimizer~\cite{DBLP:journals/corr/KingmaB14}.
Since all methods are based on CLIP, the text and frame encoders are initialized with public CLIP checkpoints (ViT-B/32 with input patch sizes of 32 and ViT-B/16 with input patch sizes of 16).
The CLIP encoders are fine-tuned with an initial learning rate of $1\mathrm{e}{-5}$, while other modules use $1\mathrm{e}{-4}$.
Learning rates are warmed up for 5 epochs then decayed to zero following a cosine schedule, with weight decay of 0.2. We train our model for 50 epochs.
For KG-CLIP, video frames follow the same sampling strategy~\cite{DBLP:journals/pami/0002X00LTG19} and data augmentation~\cite{DBLP:journals/corr/abs-2109-08472}, and are preprocessed to $224 \times 224$ spatial resolution. 
Experiments are conducted on a single NVIDIA A100-80GB GPU.

\subsection{Overall Comparison}
We conduct fully-supervised experiments on the Kinetics-TPS dataset to compare KG-CLIP against previous methods using their original setups. To comprehensively analyze the impact of the multi-modal knowledge graph, we split the data into four subsets with different train-test ratios: 2 (761) : 8 (3048), 4 (1523) : 6 (2286), 6 (2285): 4 (1524) and 8 (3047): 2 (762). We also evaluate with 4, 8, and 16 input video frames.
Results in Table \ref{tab:overall} show: 
\textbf{(1)} KG-CLIP achieves new state-of-the-art Top-1 and Top-5 accuracy across all data subsets, with significant gains over prior methods. 
\textbf{(2)} As available training data decreases, KG-CLIP shows expanding gains over baselines, with over 4\% maximum improvement, highlighting its strong multi-modal representation capabilities amid limited supervision. 
\textbf{(3)} Even with just 4 input frames, KG-CLIP produces accurate recognition that can match or surpass 16-frame baseline performance for some datasets.

We further analyze the results and try to conclude the reasons for the above phenomena. A potential reason for KG-CLIP's strong performance is that the fine-grained knowledge graph provides rich multi-modal correlations. This facilitates recognition of correct patterns even with limited training data. Without such assistance, baselines must continuously mine patterns from ample data, demanding more samples. Thus, knowledge graph-guided learning is able to improve accuracy and efficiency with limited training samples, which is also the key motivation of this research.

\subsection{Comparison with Prompt Tuning}
\begin{figure}[t]
    \centering
    \includegraphics[width=0.98\linewidth]{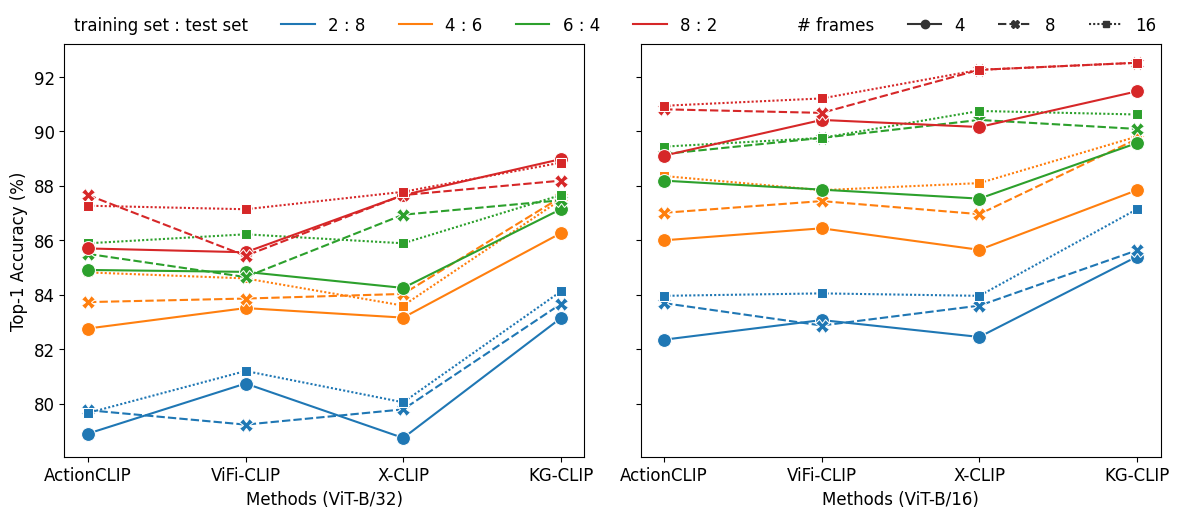}
    \caption{Comparison of KG-CLIP with baselines supported by body movement prompts on four datasets with varying numbers of video frames and backbones.}
    \label{fig:prompt}
\end{figure}

Considering that fine-grained knowledge of body movements is a newly introduced knowledge setting, we expand baselines without changing their original structure to make a fairer comparison. Specifically, we enrich the baselines' default prompts with fine-grained action descriptions of body movements, since prompt engineering is critical for CLIP-based models. For each action class, we identify body movements with the highest frequency across 10 body parts, and use these to construct detailed prompt labels. As Figure~\ref{fig:prompt} shows: \textbf{(1)} While performance of baselines improves with the addition of fine-grained knowledge in prompts, gains remain constrained as body movements are not structurally modeled. \textbf{(2)} KG-CLIP maintains superior performance over baselines, especially amid limited samples, owing to the knowledge graph's comprehensive modeling of interrelated visual and textual concepts. 

In summary, while prompt engineering does provide some performance gains for baseline methods, these improvements are limited. This further emphasizes the importance of utilizing more integrated and structured knowledge representations, rather than relying solely on surface-level descriptive prompts.

\begin{table}[t]
\centering
\caption{Ablation study results (\%) on four Kinetics-TPS datasets. For each dataset, we remove the three novel components contained in KG-CLIP one by one. 
}
\resizebox{0.98\linewidth}{!}{%
\begin{tabular}{|l|c|c|c|c|}
\hline
\multicolumn{1}{|c|}{\textbf{Backbones}} & \multicolumn{2}{c|}{\textbf{ViT-B/32}} & \multicolumn{2}{c|}{\textbf{ViT-B/16}} \\ \hline
\multicolumn{1}{|c|}{\textbf{\# frames}} & \multicolumn{2}{c|}{\textbf{8}}        & \multicolumn{2}{c|}{\textbf{8}}        \\ \hline
\multicolumn{1}{|c|}{\diagbox[dir=SE]{\textbf{Methods}}{\textbf{Accuracy}}}          & \textbf{Top-1}     & \textbf{Top-5}    & \textbf{Top-1}     & \textbf{Top-5}    \\ \hline\hline
\multicolumn{5}{|c|}{training set : test set = 2 : 8}    \\ \hline
\rowcolor[HTML]{EFEFEF} 
KG-CLIP                                  & \textbf{83.66}     & \textbf{99.25}    & \textbf{85.63}     & \textbf{99.34}    \\
-w/o multi-modal contrast                & \textbf{83.66}     & {\ul 99.21}       & 85.20        & 98.49             \\ 
-w/o reversed triplet learning           & {\ul 82.97}        & 99.05             & {\ul 85.60}              & \textbf{99.34}    \\ 
-w/o deviation compensation              & {\ul 82.97}        & 99.11             & 85.30              & {\ul 99.11}       \\ 
-w/ Euclidean distance          &     82.38               &  98.59                 &            83.86        &  98.75                 \\ \hline\hline
\multicolumn{5}{|c|}{training set : test set = 4 : 6}                                                                      \\ \hline
\rowcolor[HTML]{EFEFEF} 
KG-CLIP                                  & \textbf{87.58}     & \textbf{99.52}       & \textbf{89.72}     & \textbf{99.48}       \\
-w/o multi-modal contrast                & 87.18              & {\ul 99.48}    & 88.63              & 99.13             \\ 
-w/o reversed triplet learning           & 86.66             & 99.39             & {\ul 89.06}        & 99.34    \\ 
-w/o deviation compensation              & {\ul 87.53}        & 99.34             & 88.15              & {\ul 99.43}             \\
-w/ Euclidean distance         &    85.83                & 99.08         &   86.88          &    98.99               \\ \hline\hline
\multicolumn{5}{|c|}{training set : test set = 6 : 4}                                                                      \\ \hline
\rowcolor[HTML]{EFEFEF} 
KG-CLIP                                  & \textbf{87.47}     & \textbf{99.41}    & \textbf{90.09}     & \textbf{99.41}    \\ 
-w/o multi-modal contrast                & {\ul 87.20}        & 99.28             & 89.83              & {\ul 99.28}       \\ 
-w/o reversed triplet learning           & 86.61              & {\ul 99.34}       & {\ul 90.03}        & \textbf{99.41}             \\ 
-w/o deviation compensation              & 87.14              & 99.28             & 89.76              & 99.15             \\
-w/ Euclidean distance           &        85.70            & 99.02            &   89.57        & 98.88       \\ \hline\hline
\multicolumn{5}{|c|}{training set : test set = 8 : 2}                                                                      \\ \hline
\rowcolor[HTML]{EFEFEF} 
KG-CLIP                                  & \textbf{88.19}     & \textbf{99.21}    & \textbf{92.52}     & \textbf{99.48}    \\ 
-w/o multi-modal contrast                & 87.93              & {\ul 99.08}       & 91.34              & \textbf{99.48}    \\ 
-w/o reversed triplet learning           & {\ul 88.06}        & 98.82             & 91.99              & \textbf{99.48}    \\ 
-w/o deviation compensation              &  87.80        & 98.82    & {\ul 92.39}        & \textbf{99.48}    \\ 
-w/ Euclidean distance           &     87.14   & 98.95       &      90.55              &  98.69      \\ \hline
\end{tabular}
}
\label{tab:ablation}
\end{table}

\subsection{Ablation Study}
In this section, we conduct an ablation study to evaluate the impact of different modules in KG-CLIP. We compare the performance of KG-CLIP with and without the following components: ``multi-modal contrastive learning'' (Equation~\ref{eq:mm_loss}), ``reversed triplet learning'' (Equation~\ref{eq:tri_loss_1}), and ``deviation compensation'' (Equation~\ref{eq:dc}). 
We also examine use of a Euclidean distance metric, more common in traditional knowledge graph methods, rather than the cosine similarity measure (Equation~\ref{eq:cosine}).
Table~\ref{tab:ablation} shows the results of using 8 video frames as input.

We find that all proposed components provide notable gains in Top-1 and/or Top-5 accuracy.
We posit that these modules play complementary roles: multi-modal contrastive learning enhances the capabilities of cross-modal knowledge encoding, reversed triplets facilitate the alignments of entities and relations, deviation compensation mitigates gaps between knowledge derived from cross-modal encoding, and cosine distance measure is preferred by triplet learning in a multi-modal context.
Moreover, we observe that the effect of these components varies with the number of training samples. The more training samples are available, the more significant the improvement in Top-1 and Top-5 is. This suggests that more complex model structures can better handle large and diverse data. Regarding evaluation metrics, we take optimizing the Top-1 accuracy as the main goal of model training. Therefore, each component’s influence is predominantly observed on Top-1 accuracy. As for the Top-5 accuracy, its performance nears saturation, and hence, slight changes of the model architecture only cause fluctuations within a bounded range.

\begin{figure*}[t]
    \centering
     \subfloat[][training set : test set = 2 : 8]{\includegraphics[width=0.24\linewidth]{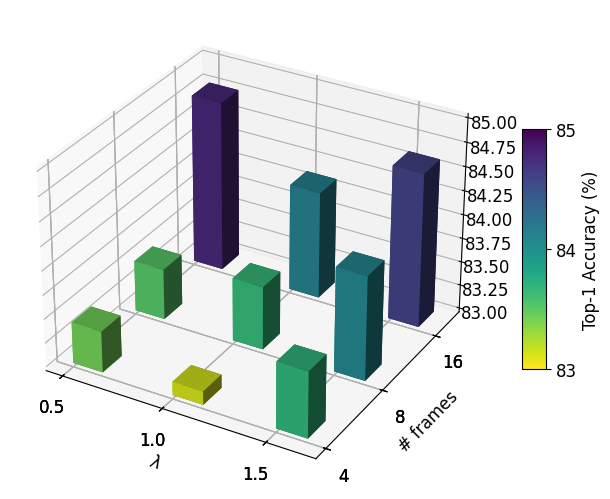}\label{fig:32_28}} 
     \hspace{2pt}
     \subfloat[][training set : test set = 4 : 6]{\includegraphics[width=0.24\linewidth]{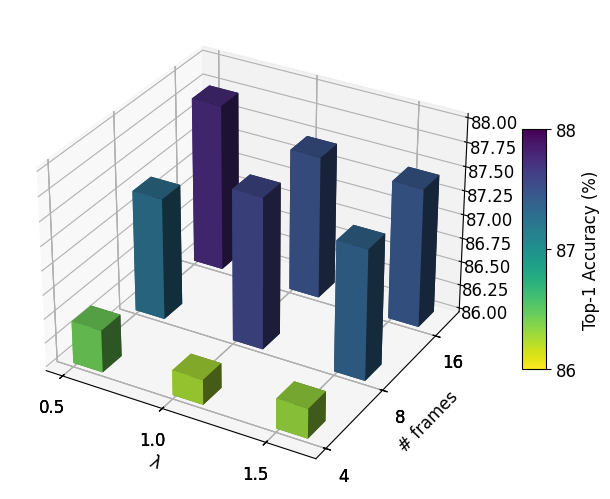}\label{fig:32_46}}
     \hspace{2pt}
     \subfloat[][training set : test set = 6 : 4]{\includegraphics[width=0.24\linewidth]{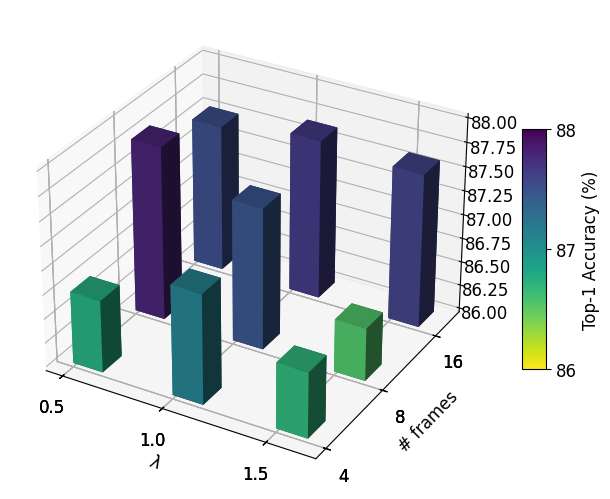}\label{fig:32_64}}
     \hspace{2pt}
     \subfloat[][training set : test set = 8 : 2]{\includegraphics[width=0.24\linewidth]{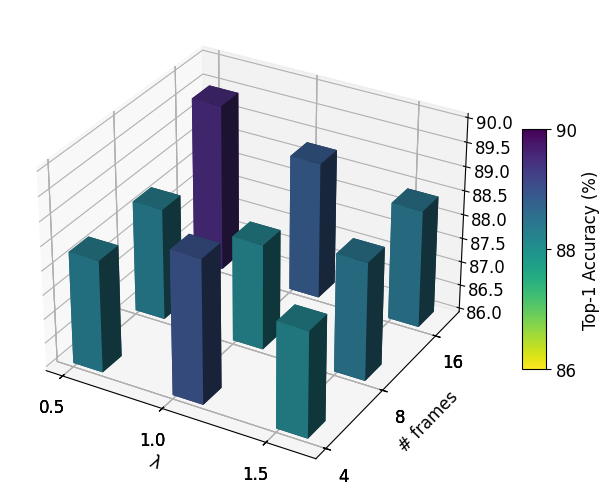}\label{fig:32_82}}
    \caption{Top-1 accuracy (\%) on four datasets with different number of video frames and multi-modal learning settings ($\lambda$) using ViT-B/32 as backbone.}
    \label{fig:lambda}
\end{figure*}
\subsection{Parameter Analysis}
Through our ablation studies, we have discerned the important role of multi-modal learning. Despite its relevance being confined to two types of multi-modal relationships, multi-modal learning exerts a substantial influence on global optimization. Consequently, we have undertaken a detailed examination of the performance sensitivity concerning the multi-modal loss coefficient, denoted as $\lambda$ in Equation~\ref{eq:final_loss}. 
Our experimental methodology involves varying the value of $\lambda$ and the number of input frames. The outcomes of these experiments are presented in Figure~\ref{fig:lambda}, which is conducted on the four datasets with distinct segmentation ratios and uses ViT-B/32 as the backbone.
The results reveal that the optimal $\lambda$ value varies depending on the dataset and the number of input frames. However, a consistent observation across most combinations is that a $\lambda$ value of 1.0 can lead to more stable performance, making it a relatively optimal choice across all combinations. This insight can guide parameter tuning when processing new data or selecting different numbers of video frames for learning, e.g., initiating the model with a multi-modal learning coefficient of 1.0.

\subsection{Model Efficiency}

We calculate Floating Point Operations (FLOPs) and the number of parameters of each model for model efficiency analysis. It should be noted that all training parameters are counted in this experiments, and the light-weight core library fvcore\footnote{https://github.com/facebookresearch/fvcore/tree/main\label{fn:flops}} is adopted for the calculation of FLOPs.
As shown in Table~\ref{tab:efficiency}, with ViT-B/32 as the backbone network, the FLOPs of our proposed KG-CLIP is slightly higher than ActionCLIP, but significantly lower than X-CLIP and ViFi-CLIP. But since our model encodes three types of triplets in parallel, the FLOPs exceeds all baselines when using ViT-B/16.
It’s important to note that ViT-B/16 utilizes smaller patches, resulting in higher FLOPs and relatively better accuracy, while ViT-B/32, despite having fewer FLOPs and faster inference speed, lags slightly in accuracy compared to ViT-B/16. 
Under this trade-off, many contemporary models prioritize larger pre-trained models, such as ViT-B/16 or even ViT-Large models, in pursuit of higher accuracy, which often comes at the cost of training and inference efficiency. In contrast, our proposed KG-CLIP model offers a good solution for this challenge. Despite a marginal increase in FLOPs and parameters compared to the existing baselines, KG-CLIP can achieve superior performance only using a limited number of video frames and a basic ViT-B/32 backbone network. 
Remarkably, with just four frames, KG-CLIP is able to outperform some baselines that rely on 16 frames, which also underscores the key advantage of our method.
\begin{table}[t]
\caption{Model efficiency analysis on FLOPs and number of parameters based on 8 video frames per video as visual inputs.}
\begin{tabular}{|ccc|}
\hline
\multicolumn{1}{|c|}{\textbf{Methods}}                & \multicolumn{1}{c|}{\textbf{FLOPs (G)}}       & \textbf{\# parameters (M)} \\ \hline
\multicolumn{3}{|c|}{Backbone: ViT-B/32}                                                                                       \\ \hline
\multicolumn{1}{|c|}{ActionCLIP}                      & \multicolumn{1}{c|}{38.41}                    & 170.2                  \\ 
\multicolumn{1}{|c|}{ViFi-CLIP}                       & \multicolumn{1}{c|}{106.07}                   & 126                    \\ 
\multicolumn{1}{|c|}{X-CLIP}                          & \multicolumn{1}{c|}{106.96}                   & 196.6                  \\ 
\rowcolor[HTML]{EFEFEF} 
\multicolumn{1}{|c|}{\cellcolor[HTML]{EFEFEF}KG-CLIP} & \multicolumn{1}{c|}{\cellcolor[HTML]{EFEFEF}81.75} &  160.74               \\ \hline\hline
\multicolumn{3}{|c|}{Backbone: ViT-B/16}                                                                                       \\ \hline
\multicolumn{1}{|c|}{ActionCLIP}                      & \multicolumn{1}{c|}{143.75}                   & 168.6                  \\ 
\multicolumn{1}{|c|}{ViFi-CLIP}                       & \multicolumn{1}{c|}{211.42}                   & 124.3                  \\ 
\multicolumn{1}{|c|}{X-CLIP}                          & \multicolumn{1}{c|}{212.98}                   & 194.9                  \\ 
\rowcolor[HTML]{EFEFEF} 
\multicolumn{1}{|c|}{\cellcolor[HTML]{EFEFEF}KG-CLIP} & \multicolumn{1}{c|}{\cellcolor[HTML]{EFEFEF}281.75} & 159.08                      \\ \hline
\end{tabular}
\label{tab:efficiency}
\end{table}
\section{Conclusion}
\label{sec:con}
In this paper, we propose a novel video-language learning framework, KG-CLIP, to enhance video action recognition by injecting a knowledge graph based on fine-grained action parsing. We construct a multi-modal knowledge graph composed of multi-grained concepts by action parsing, and innovatively design a triplet learning component to model multi-relation knowledge graphs and map entities into a shared multi-modal space. In particular, to address the modality gap problem in joint representation across vision and language, we implement a deviation compensation technique to actively bridge the knowledge gap, thereby improving knowledge alignment to enhance model performance. 
Comprehensive experiments on the large-scale Kinetics-TPS dataset demonstrate improved recognition over state-of-the-art methods. Moreover, KG-CLIP exhibits excellent sample efficiency and learning capacity amid limited input frames or training data. Given the marked performance improvements across settings combined with acceptable efficiency overheads, this work highlights the viability of structured knowledge graph representations for guiding vision-language models to better video understanding.


\bibliographystyle{ACM-Reference-Format}
\bibliography{sample-base}

\end{document}